# Practical way to increase nonlinearity of kinetic inductance of superconductor

Korneeva Yu. P.[1], Dryazgov M. A.[1], Trofimov I. V.[1], Levichev M. Yu.[2], Porokhov N. V.[1], Mumlyakov A. M.[1], Shibalov M. V.[1], Vodolazov D. Yu.[1,2], Korneev A. A.[1,3], Tarkhov M. A.[1]

[1]Institute of Nanotechnology of Microelectronics of the Russian Academy of Sciences, 115487, Moscow, Nagatinskaya 16A, build. 11, Russia.

[2]Institute for Physics of Microstructures, Russian Academy of Sciences, 603950, Nizhny Novgorod, GSP-105, Russia

[3]Higher School of Economics - National Research University, 101000, Moscow, Myasnitskaya 20, Russia

E-mail: korneeva_yuliya@mail.ru

## Abstract

This work demonstrates that depositing a thin layer of Mo (5-15 nm) onto a 10 nm thick NbN strip leads to a significant increase in the nonlinearity of the kinetic inductance $L_k$. Specifically, the change in $L_k$ with increasing current reached 70% in the NbN/Mo bilayer at liquid helium temperature, whereas in the NbN strip, $L_k$ changed by only 10% in the superconducting state. In addition to altering the nonlinear properties, the Mo layer caused a significant increase in the critical current at low temperatures (up to 2 times in the case of a 5 nm thick Mo layer). The increased nonlinearity of $L_k$ can be explained by two factors: i) a reduction of the critical supervelocity at which the superconducting state becomes unstable with respect to vortex formation when a Mo layer is deposited on NbN, and ii) a higher sensitivity of the induced superconductivity in Mo to supervelocity/supercurrent. Considering the results on the transport properties of SN bilayers with a high ratio of layer resistivities $\rho_S/\rho_N \gg 1$, it can be concluded that depositing a thin layer of a relatively low-resistivity metal onto a superconductor with high $\rho$ is a practical method for achieving a large nonlinearity of the superconductor's kinetic inductance.

## 1. Introduction

In superconductors, in addition to the geometric inductance $L_g$ (associated with the energy of the magnetic field generated by the current), there exists kinetic inductance $L_k$, which is determined by the kinetic energy of the moving superconducting electrons (in normal metals, this contribution is usually negligible up to frequencies on the order of $1/\tau$, where $\tau$ is the electron mean free time). Unlike $L_g$, the kinetic inductance depends on the current ($L_k$ increases with the increase of the current), which is associated with the reduction in the number of superconducting electrons as the current grows. This property of kinetic inductance has been proposed for use in various applications: 1) parametric amplifiers, where the effect of amplifying a weak signal/current arises from its mixing with a reference signal of sufficiently larger amplitude [1-3]; 2) magnetometers in the form of a superconducting ring, in which an external magnetic field induces a shielding current that increases the kinetic inductance, allowing its change to be measured in an LC circuit by the shift in its resonant frequency and the change in quality factor [4]; 3) qubits operating in the regime of an anharmonic oscillator [5, 6]; 4) current sensors [7], 5) resonators with current [8] or magnetic field [9], tunable resonant frequency [10]; 6) current controlled variable delay superconducting transmission line [11].

It is evident that the greater the change in $L_k$ with varying current, the better it is for applications. In a conventional superconductor, a significant change (high nonlinearity) in the kinetic inductance is achieved when the current approaches the depairing current. For instance, figure 1 (red curves) shows the dependence of the supercurrent $I$ on the supervelocity $q$ ($q = \nabla\theta$, where $\theta$ is the phase of the superconducting order parameter), which follows from the Usadel model for a superconducting strip under the condition of a uniform current distribution across its width (this condition is met when the strip width $w \lesssim \lambda^2/d$, where $\lambda$ is the magnetic field penetration depth in the superconductor, $d$ is the strip thickness, and the contribution of the vector potential to $q$ from the current's magnetic field can be neglected). The inset in figure 1 illustrates the dependence $L_k \sim dq/dI$. When the current reaches $I_{dep,S}/2$, the kinetic inductance increases by only about 10%, whereas $L_k$ diverges as the current approaches $I_{dep,S}$. In real superconductors, it is difficult to reach the depairing current due to inhomogeneities (edge roughness or variations in material composition). This is because the supervelocity locally reaches its critical value near an inhomogeneity at a current lower than $I_{dep,S}$, causing the formation of vortices in the strip [12]. Vortex motion under the influence of the current heats the superconductor, leading to its transition to the normal state. Therefore, in real experiments, the change in $L_k$ typically amounts to a few percent [4, 7, 10, 13, 14].

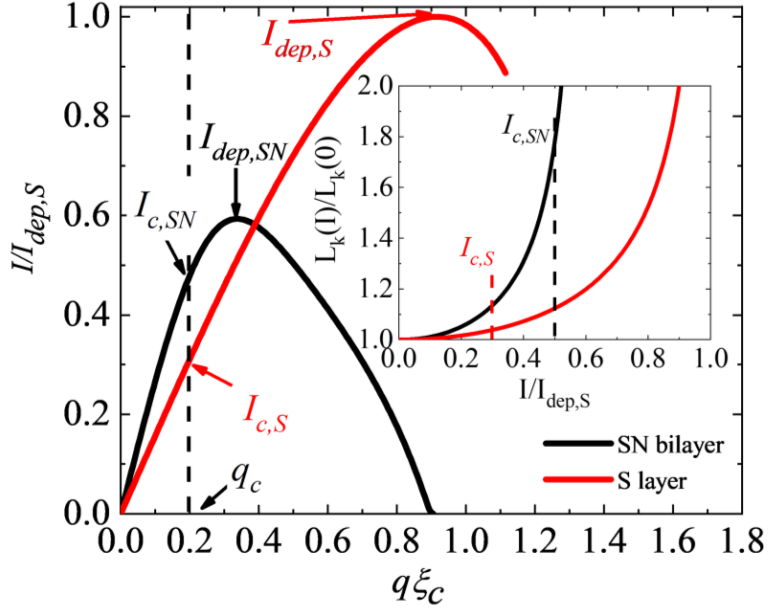

**Figure 1**. Dependence of the supercurrent $I$ on the supervelocity $q\xi_C$ and the kinetic inductance $L_k = \frac{l\hbar}{2e} \cdot \frac{dq}{dI}$ on the current (in the inset) for a conventional superconductor (red) and an SN-bilayer (black). The temperature is chosen at $T = 0.1 T_{c0}$ (where $T_{c0}$ is the critical temperature of the superconductor, $l$ is the length of the superconductor), the ratio of diffusion coefficients $D_N/D_S = 10$, the density of states at the Fermi level is assumed to be the same for both the S and N layers, and the thicknesses of the superconductor and normal metal are $d_S = 1.5\xi_C$ and $d_N = 0.75\xi_C$, respectively, where $\xi_C = \left(\frac{\hbar D_S}{k_B T_{c0}}\right)^{1/2}$. The figure illustrates the possible reason for the increase in the nonlinearity of the superconductor's kinetic inductance and the critical current when a thin layer of a low-resistivity normal metal is deposited on it. When the critical momentum value remains unchanged upon the addition of the normal metal layer (in the figure, $q_c \xi_c = 0.2$), this leads to an increase in the critical current from $I_{c,S}$ to $I_{c,SN}$ and a significantly stronger change in $L_k$ (see the inset).

In a recent study [15], it was demonstrated that coating a superconducting MoN strip with a thin layer of copper results in a nonlinear dependence of the critical current on the magnetic field in weak fields. This indicates that the critical current approaches the depairing current of such a bilayer superconductor – normal metal (SN) strip [15], and consequently, one can expect an increase in the nonlinearity of $L_k(I)$ in such a structure. Additionally, it was found that the critical current in zero magnetic field also increases. Both of these effects can be explained if we assume that superconductivity breaks down at the same critical supervelocity (or slightly lower) for both a conventional S-strip and an SN-strip. Figure 1 illustrates this effect. It shows the $I(q)$ dependencies for S- (red curves) and SN-strips (black). At fixed chosen $q_c \xi_c = 0.2$, the addition of an N-layer increases the critical current, while simultaneously causing a significantly larger change in $L_k$ (see the inset in figure 1). Note that this effect does not occur under arbitrary parameters but rather under the condition that the electron diffusion coefficient in the normal metal $D_N$ is significantly larger than $D_S$ (or equivalently, when $\rho_N/\rho_S \ll 1$). This condition is necessary for a sufficiently large supercurrent to flow through the N-layer when superconductivity is induced in it, as well as for its high sensitivity to supervelocity due to the larger value of $D_N$ [15]. As seen in figure 1, at $q_c \xi_c = 0.2$, we are significantly closer to the depairing current of the SN-bilayer than to that of a single superconductor, which leads to a stronger change in $L_k$.

The condition that supervelocity reaches its critical value, rather than the current at which the superconducting state breaks down due to the appearance of vortices, has been discussed in [12, 16] for conventional superconductors. For SN-bilayers, such studies have not been conducted, and this is our assumption, which qualitatively explains the results of the study [15].

In this work, we experimentally investigate the transport characteristics of SN-strips consisting of a superconducting NbN layer approximately 10-nm-thick and a Mo layer of varying thickness. Mo is a superconductor with a critical temperature of about 0.9 K (and can be considered a normal metal at higher temperatures) and has a significantly lower normal-state resistivity compared to NbN. Unlike the study [15], we use different superconducting and normal materials, thinner metallic layers, and directly measure the $L_k(I)$ dependencies to address the question of changes in the nonlinearity of $L_k(I)$, without relying on measurements of the critical current as a function of the magnetic field. We found that coating NbN with a Mo layer significantly increases the critical current at low temperatures despite the reduction in the critical temperature. Our direct measurements of the $L_k(I)$ dependence at liquid helium temperature (4.2 K) showed that in the NbN/Mo system, a 70% change in kinetic inductance can be achieved, whereas in NbN, the kinetic inductance changed by only 10%. This result is the main finding of our work. Thus, we have directly demonstrated that using a thin layer of a normal metal (or a superconductor with a low $T_C$) with a much lower resistivity than that of the superconductor in the normal state can significantly enhance the nonlinearity of the kinetic inductance. This effect can be achieved for sufficiently thin films with a large kinetic inductance (for a hybrid of 10-nm-thick NbN and 5-nm-thick Mo, $L_k$ at $I = 0$ is approximately 30 pH per square at 4.2 K).

## 2. Experimental details

### A. Samples fabrication

The samples were fabricated using thin-film planar technology. Low-resistivity silicon wafers, pre-oxidized at 1000°C in a wet oxygen atmosphere, were used as substrates. The bilayer NbN-Mo system was deposited using DC magnetron sputtering in an in-situ process onto substrates preheated to 350°C. First, the NbN film was deposited in a mixture of Ar and N₂ gases, followed by the deposition of the Mo film in an Ar atmosphere without breaking the vacuum (*in situ*). The thickness of the superconducting layer $d_{NbN}$ was approximately 10 nm for all samples, while the thickness of the molybdenum layer $d_{Mo}$ varied in the range of 5 nm to 14 nm. The exact thicknesses of the layers were measured using X-ray reflectometry (XRR) and are presented in Table 1. For comparison, NbN films without molybdenum (denoted as NbN in Table 1) were also fabricated.

**Table 1.** Sample characteristics.

| Sample | $d_{NbN}$ (nm) | $d_{Mo}$ (nm) | $R_s$(300K) ($\Omega/\square$) | $T_c$ (K) | $L_k(0)$ (nH) | $L_s$ (pH/$\square$) | $R_{300K}/R_{20K}$ | $\rho$ ($\mu\Omega$*cm) |
|---|---|---|---|---|---|---|---|---|
| NbN | 9.6 | --- | 416 | 8.5 | 55.7 | 80.9 | 0.81 | 399.4 |
| SN5 | 10.4 | 4.9 | 82 | 6.9 | 21.6 | 35.2 | 0.98 | 40.2 |
| SN6 | 10.2 | 9.6 | 42 | 6.8 | 17.7 | 30.3 | 1.1 | 40.3 |
| SN7 | 10.2 | 14.2 | 25 | 6.6 | 16.9 | 29.2 | 1.14 | 35.5 |

Straight strips with a width of 4 μm and a length of 3 mm were fabricated using optical laser lithography and plasma-chemical etching. To prevent oxidation, the NbN-Mo bilayer film was coated with a protective $SiN_x$ layer with a thickness of 12 nm. An optical photograph of the sample is shown in figure 2.

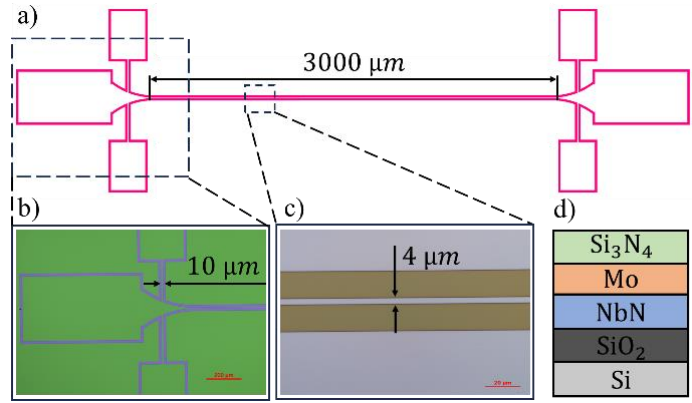

**Figure 2.** (a) Sample topology (not to scale) with optical photographs of sections (b-c); (d) layer sequence.

### B. Measurement methods

The resistances of the studied samples were measured using a four-probe method with a Keithley DMM6500 digital multimeter at room temperature. From these resistances and the geometric dimensions of the samples, the sheet resistances $R_s$ (resistances per square of the film) as well as the resistivities $\rho$ were calculated, as presented in Table 1.

Measurements of the temperature dependence of resistance and the temperature dependence of the critical current of the samples were carried out in a Gifford-McMahon cryocooler, which allows reaching a minimum temperature of 2.6 K. Resistance measurements in the temperature range of 2.6 K to 300 K were performed during the cooling/heating process of the cryocooler.

When measuring the critical temperature and the dependencies of the critical current on temperature, heating/cooling in the range of 2.5 K to 15 K was carried out using a heater installed on the cold plate of the cryostat, controlled by a LakeShore 325 temperature controller. For measuring $R(T)$ of the samples, a Keithley SourceMeter 2460 was used, connected in a two-probe configuration. To reduce noise, a current of 10 µA was supplied to the sample through the DC port of a bias tee MiniCircuits ZFBT-282-1.5A+. The critical temperature $T_C$ was defined as the temperature at which the resistance equals half of $R_{20}$ (resistance at 20 K temperature).

To measure the critical current as a function of temperature at a given temperature, a set of 10 current-voltage characteristics (IV curves) was recorded over approximately one minute using a source-measure unit in voltage source mode. The critical current was selected as the maximum of the critical currents measured from these ten IV curves. This procedure allowed us to account for small temperature oscillations inherent to the Gifford-McMahon cycle of the cryocooler.

Measurements of the dependence of kinetic inductance on current were carried out in liquid helium inside a transport Dewar at a temperature of 4.2 K. In the experiment, the impedance of NbN and NbN/Mo strips was measured as a function of the DC current through the superconducting strip using a lock-in amplifier (Stanford Research SR830). The imaginary part of the impedance is proportional to $2\pi f(L_k + L_g)$, while the real part is proportional to the resistance of the strip $R$ (before reaching the critical current, $R = 0$). Measurements were performed at a frequency of $f =$ 100 kHz with an AC current amplitude of 50 µA, and the current through the strip was controlled using a current source. A four-point probe configuration was used in the measurements. For strips of our size, using the known expression for the geometric inductance of a strip [17], we obtained $L_g \sim$ 5 nH. This value was subtracted from the total inductance $L_k + L_g$ obtained in the measurements.

## 3. Results and discussion

Table 1 presents the thicknesses $d$ of the studied samples, their sheet resistances $R_s$, and resistivities $\rho$, measured at room temperature. The ratio of resistivities $\rho_{NbN}/\rho_{Mo}$ for all Mo thicknesses was approximately 10. The choice of the NbN layer thickness is determined by several factors. It should not be too large, otherwise the kinetic inductance will be small (since $L_k \sim 1/d$), and the contribution of the normal metal to the transport characteristics ($L_k$ and $I_c$) will be negligible, preventing a significant increase in the nonlinearity of $L_k(I)$. On the other hand, if the superconductor thickness is too small, the critical temperature of the bilayer may decrease significantly or even drop to zero due to the inverse proximity effect. The optimal thickness is on the order of 1-3

$\xi_s(0)$ (where $\xi_s(0)$ is the coherence length at zero temperature, which is approximately 5 nm for NbN). We chose a thickness in the middle of this range: $d_{NbN} \sim 10$ nm $\sim 2\xi_s(0)$.

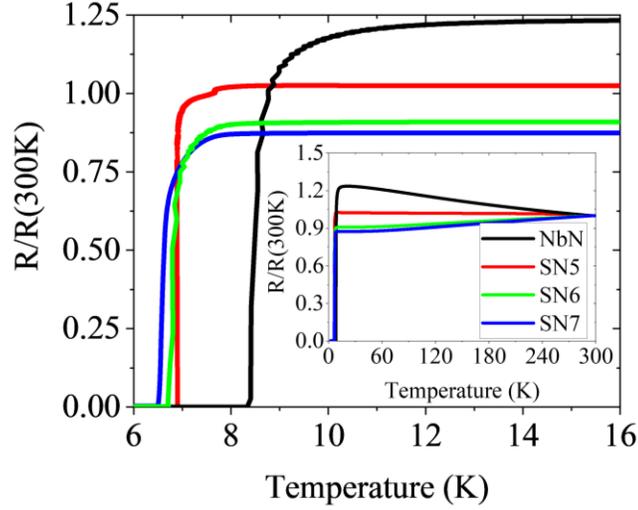

**Figure 3**. Temperature dependencies of resistance. Increasing the thickness of the normal metal layer leads to a decrease in the critical temperature, which is a sign of the proximity effect. The temperature dependence of resistance for the reference sample (NbN without Mo) is typical for NbN.

Figure 3 shows the temperature dependencies of resistance. The critical temperature of the reference sample without molybdenum (NbN) was 8.5 K. The residual resistivity ratio (RRR), defined as the ratio of the resistance at room temperature $R_{300}$ to the resistance at 20 K ($R_{20}$), ranges from 0.81 for pure NbN without Mo to 1.14 for the sample with the maximum Mo thickness. The Mo layer, with its lower resistivity, shunts the high-resistivity NbN. The critical temperature of the SN samples did not exceed 7 K, which is a consequence of the inverse proximity effect.

Figure 4 shows the temperature dependencies of $I_c(T)$ for NbN/Mo samples (red, green, and blue curves) and for the NbN sample (black curve). The measured temperature dependence $I_c(T)$ is best approximated by the Ginzburg-Landau expression for the depairing current $I_{dep}(T) = I_{dep}(0)(1 - t^2)^{3/2}(1 + t^2)^{1/2}$, where $t = T/T_C$ [18], which is applicable to the two-fluid model at all temperatures [19]. The theoretical curve for $I_{dep}(T)/I_{dep}(0)$ is also shown on the graph as a black dashed curve. In the calculations, $I_{dep}(0) = 0.74w(\Delta(0))^{3/2}/eR_s\sqrt{\hbar D}$ was used (according to our estimates, $I_{dep}(0) = 1.5$ mA, using $D = 0.5$ cm²/s, $R_s$ and $T_c$ are taken from Table 1), where $\Delta(0) \approx 1{,}76 k_B T_c$.

The critical current of the S sample is 60% of the theoretical depairing current. When a Mo layer is deposited, the critical current of the SN-sample doubles at T=2.5 K for a molybdenum thickness of $d_{Mo}$=4.9 nm.

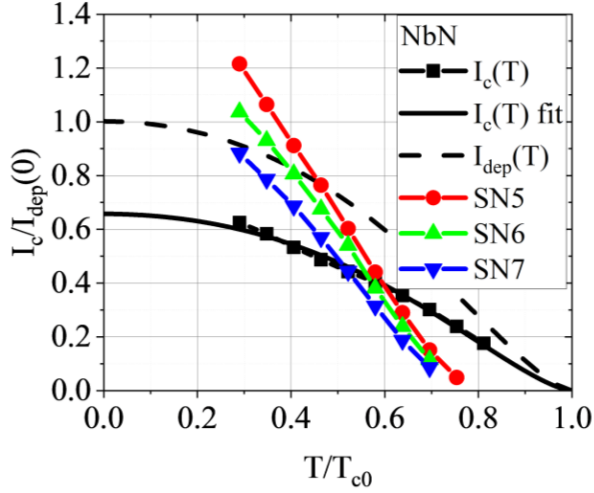

**Figure 4**. Dependence of $I_c/I_{dep}$ on temperature.

The presence of a Mo layer leads to an increase in the critical current at sufficiently low temperatures, and the thinner the Mo layer, the higher the critical current. A similar effect of increasing $I_c$ has been observed for the Cu/MoN pair [15], NbN/Al and NbN/Ag pairs [20], as well as the NbN/CuNi pair [21, 22]. In all cases, the normal metal layer (or, in the case of Al, a superconductor with a lower $T_c$) had a significantly lower resistivity than the superconductor. Thus, we arrive at a situation qualitatively depicted in figure 1, where the normal layer contributes significantly to the supercurrent, and at a fixed supervelocity, the supercurrent is higher in the SN bilayer than in a single S-layer. This leads to an increase in the critical current, assuming that the critical supervelocity value remains unchanged. Note that even if $q_c$ decreases in the bilayer, the critical current may still be higher than in a single S-layer due to the steeper slope of $I(q)$ at small $q$ (see figure 1).

In addition to the increase in $I_c$, the temperature dependence of $I_c(T)$ also becomes steeper. This effect is associated with the smaller value of the energy gap $\varepsilon_g$ in the electron spectrum of the SN bilayer compared to the superconductor. As a result, the saturation of the $I_c(T)$ dependence occurs at a lower temperature $T \lesssim T_g = \varepsilon_g/k_B$ (the thicker the normal metal layer, the smaller $\varepsilon_g \sim 1/d_N^2$ [23], and consequently, $T_g$).

The dependence of $\varepsilon_g$ on the thickness of the normal metal layer also qualitatively explains the higher value of $I_c$ for the bilayer with a smaller thickness of the normal layer, assuming the independence of $\rho_N$ on $d_N$. Indeed, the supercurrent $I_N$ flowing through the normal layer is proportional to the square of the induced superconducting order parameter $\Delta$: $I_N \sim d_N \Delta^2/\rho_N$ (at temperatures $T < T_g$, $I_N \sim d_N \Delta/\rho_N$), as follows from the Usadel model, which is valid for 'dirty' superconductors. Considering that $\Delta \sim \varepsilon_g \sim 1/d_N^2$, this leads to a smaller $I_c$ for larger $d_N$, despite the increase in the thickness of the N layer. An illustration of this effect is shown in Figure 4b of [20], where the calculated dependencies of the depairing current $I_{dep,SN}$ on temperature for the SN bilayer are presented for various $d_N$ and a fixed $\rho_N$.

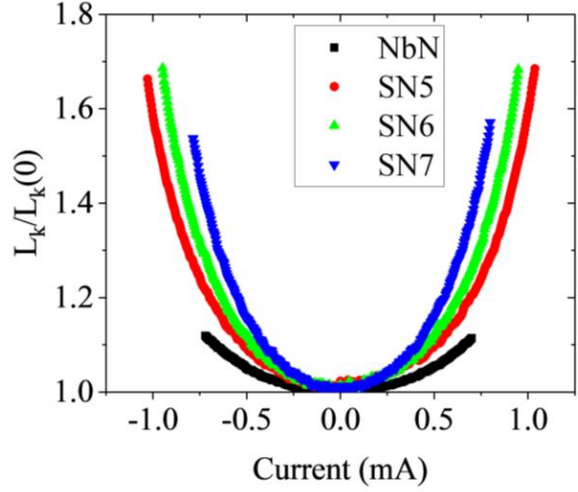

**Figure 5**. Dependencies of kinetic inductance on current for various samples, measured at T = 4.2 K.

Figure 5 shows the measured dependencies of kinetic inductance on current for NbN and NbN/Mo bilayers. The maximum increase in $L_k$ by 10% at a critical current $I_c \sim 0.6\, I_{dep}$ for the NbN strip correlates with the result obtained from the Usadel model (although the inset in figure 1 shows the $L_k(I)$ dependence at $T = 0.1\, T_{c0}$, in the presented coordinates, it changes only slightly up to $0.9\, T_{c0}$). The presence of the Mo layer leads to a significantly larger relative change in kinetic inductance of about 70%, the possibility of which is qualitatively explained in figure 1.

For a quantitative description of the obtained results, it is necessary to know the ratio of diffusion coefficients $D_S/D_N$, the densities of states at the Fermi level $N_N/N_S$, as well as the barrier for electron transmission between the S and N layers $\gamma = R_{SN} A_{SN}/\rho_N$ (where $R_{SN}$ is the resistance of the SN interface, and $A_{SN}$ is its area). We only know the ratio of resistivities $\rho_S/\rho_N = D_N N_N/D_S N_S \sim 10$. To demonstrate the influence of parameter variations on the results, figure 6 shows the calculated $I(q)$ dependencies at $T = 0.5 T_{c0} \approx 4.2$ K for an S-layer with thickness $d_N = 1.5\, \xi_c$ and an SN-layer with thicknesses $d_S = 1.5\, \xi_c$, $d_N = 0.75\, \xi_c$ (where $\xi_c = \left(\frac{\hbar D_S}{k_B T_{c0}}\right)^{1/2} \approx 6.7$ nm for NbN with $D_S = 0.5$ cm²/s and $T_{c0} = 8.5$ K) and various parameters for which $\rho_S/\rho_N = 10$. In the calculations, we considered that the critical temperature of Mo $T_{c,Mo} = 0.9\text{K} \sim 0.11 T_{c0}$, which quantitatively has a significant influence even at T = 4.2. The Usadel model was used for the calculations (the equations and calculation method are described in [20]).

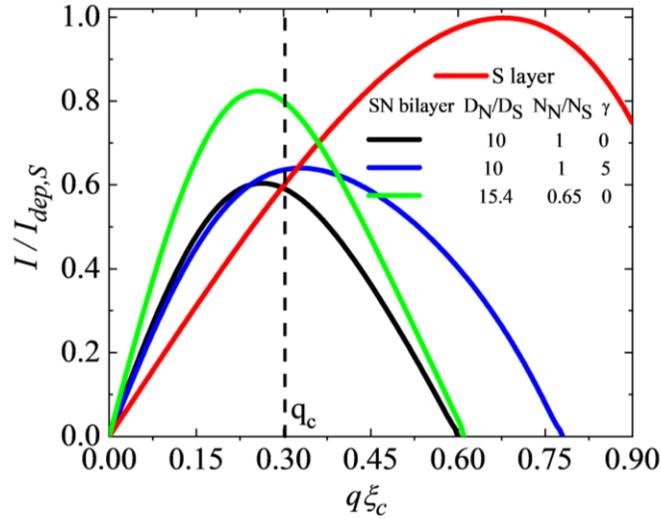

**Figure 6**. Calculated, using the Usadel model, dependence of the supercurrent on supervelocity for SN bilayers with thicknesses $d_S = 1.5\ \xi_c$, $d_N = 0.75\ \xi_c$, and a single superconductor with $d_S = 1.5\ \xi_c$ at $T = T_{c0}/2$. The remaining parameters of the model are provided in the figure.

Due to the large diffusion coefficient in the Mo layer, the depairing effect of supervelocity becomes stronger at smaller $q$, leading to a shift in the maximum of the $I(q)$ dependence compared to a single superconductor. The finite barrier value, controlled by the parameter $\gamma$, reduces the inverse proximity effect, which enhances superconductivity in the S-layer and results in a slight increase in $I_{dep,SN}$ and a shift of the $I(q)$ maximum to larger $q$. At the same time, a decrease in the ratio $N_N/N_S$ and a simultaneous increase in the ratio $D_N/D_S$ lead to an increase in the supercurrent flowing through the N layer and $I_{dep,SN}$.

Considering that for NbN $I_c(4.2K) \sim 0.6 I_{dep,S}$, we found the critical velocity $q_c \xi_c \sim 0.3$ (vertical line in figure 6). It can be seen that at this value, the supervelocity either exceeds the maximum possible value where the derivative $dq/dI \sim L_k$ for SN-bilayers remains positive (only in this case does the state with a given current remain stable) or approaches it. This means that most likely, in the SN bilayer, $q_c$ is smaller than in a single superconductor. However, it cannot be significantly smaller, as this would not lead to an increase in the critical current and the nonlinearity of $L_k(I)$.

Thus, the calculations confirm the mechanism of increasing the critical current and the nonlinearity of $L_k$. However, the lack of necessary parameters prevents a quantitative comparison between the experiment and theory.

## 4. Conclusion

We have demonstrated that even a thin (5 nm) layer of Mo deposited on a thin (10 nm) superconducting NbN strip can significantly increase the nonlinearity of its kinetic inductance. The effect is explained by the substantial contribution of Mo to the transport properties of the bilayer (for example, the kinetic inductance decreases by a factor of two compared to the NbN strip at T = 4.2) and its greater sensitivity to supervelocity/current due to the smaller value of the diffusion coefficient. Based on our results, as well as results from other SN bilayers with a large resistivity ratio, we can propose a practical method for increasing the nonlinearity of kinetic inductance. Specifically, coating a superconductor with a thin layer of a normal metal with an order of magnitude lower resistivity can lead to a relatively small decrease in the critical temperature and kinetic inductance, but a significant increase in the nonlinearity of the kinetic inductance.

## 5. Acknowledgements

The study was supported by project No. 125020501540-9 of the Ministry of Education and Science of the Russian Federation. Fabrication and technology characterization were carried out at large scale facility complex for heterogeneous integration technologies and silicon+carbon nanotechnologies.